# Computational Psychiatry in Borderline Personality Disorder

Sarah K Fineberg MD PhD*[1], Dylan Stahl[1,2], Philip Corlett PhD[1]

1 Yale University Department of Psychiatry

2 Knox College

* Corresponding author

Corresponding author:
Sarah Kathryn Fineberg
Connecticut Mental Health Center Room 518
34 Park Street
New Haven Connecticut 06519
Sarah.fineberg@yale.edu

Sources of support:
SKF is supported by a NARSAD Young Investigator Grant from the Brain and Behavior Research Foundation and by the Connecticut Mental Health Center. DSS was supported by the Richter Memorial Fund and the NSF COAST Award. PRC was funded by an IMHRO / Janssen Rising Star Translational Research Award and CTSA Grant Number UL1 TR000142 from the National Center for Research Resources (NCRR) and the National Center for Advancing Translational Science (NCATS), components of the National Institutes of Health (NIH), and NIH roadmap for Medical Research. The contents of this work are solely the responsibility of the authors and do not necessarily represent the official view of NIH or the CMHC/DMHAS.

This paper is available in final form at: https://link.springer.com/article/10.1007/s40473-017-0104-y

**Abstract:**

**Purpose of review:** *We review the literature on the use and potential use of computational psychiatry methods in Borderline Personality Disorder.*

**Recent findings:** *Computational approaches have been used in psychiatry to increase our understanding of the molecular, circuit, and behavioral basis of mental illness. This is of particular interest in BPD, where the collection of ecologically valid data, especially in interpersonal settings, is becoming more common and more often subject to quantification. Methods that test learning and memory in social contexts, collect data from real-world settings, and relate behavior to molecular and circuit networks are yielding data of particular interest.*

**Summary:** *Research in BPD should focus on collaborative efforts to design and interpret experiments with direct relevance to core BPD symptoms and potential for translation to the clinic.*

**Paper DRAFT:**
# I. Introduction

A rich literature on borderline personality disorder (BPD) has grown from clinical observation, follow-up studies, and behavioral experiments, but little is available in terms of a definitive molecular or circuit-based understanding of the disorder. Novel approaches to experimental design and data analysis may help us to better link clinical problems with laboratory based behavioral results and specific biological mechanisms. In recent years, the growth of computational power and biologically-informed mathematical models of cognition and the brain have encouraged the development of computational psychiatry. This method interprets findings and makes new predictions from behavioral and neural datasets by applying experimental observations to biologically-informed computational models [1]. This approach not only increases mechanistic understanding, but it also holds promise for improving diagnostic categories and personalizing treatment approaches [1], [2].

Despite the lack of a firm biologically-based theory, BPD researchers have been prescient in gathering ecologically-valid data sets using state of the art sensor and virtual reality based approaches. Applying these data to computational models may be a particularly good fit in this field of psychiatry [3].

## II. Computational Psychiatry

Computational psychiatry entails a variety of approaches. Common to all of these is a model, often expressed as equations, that distills some feature of mental illness. These models aim to parse complex data sets, giving insight to problems like how psychological processes might be instantiated in neural machinery or how to use patterns in large data sets to make prognostic inferences[4] [5] [6] [7]. We review several examples of computational work with relevance to BPD.

*II.A. Behavioral tasks to query specific aspects of cognition*
One method is to conduct behavioral experiments in which subjects play multi-trial games designed to query specific aspects of perception, learning and memory. Trial-by-trial subject responses are fit to computational models, yielding mathematical estimates of subjects' decision processes (reviewed in [2]). Electrophysiology, EEG, and fMRI data have been linked to these parameter estimates so that trial-by-trial behavioral responses can ultimately be understood in terms of brain regions and chemicals. Of particular interest, differences in model parameters between groups can help to arbitrate between alternative hypotheses about cognitive processes, for example how a patient group may differ from controls, or how a pharmacologic intervention differs from placebo.

One multi-trial game, the social valuation task (SVT), involves binary choices based on social and non-social cues on each of 100+ trials [8]. The SVT is a decision-making game wherein participants weigh two types of information in making their choices. First, they consider their own experiences of reward contingencies (e.g. between the color of task stimuli and reward availability). On the other hand, they meet and interact with another player before the game, and see advice (supposedly from that player, really from the computer) about which choice to make on each trial. They know that the other player's goals may not be aligned with their own. Importantly the probability (and stability of probability) of reward and color and advice vary separately over the course of the game. Non-psychiatric control subjects develop beliefs as they play about how much to trust the colors and the advice.

A recent exciting development is the application of this kind of computer game to social cognition across dimensions of social symptoms. In a recent collaborative effort between computational and clinical researchers, Sevgi et al. used a game very similar to the SVT to quantify relative attention to social and non-social cues, and validated the computational model by showing that this parameter (relative weighting of social/non-social cues) linearly decreases with increasing autism-like symptoms in the general population [9]. We have hypothesized that BPD is maintained by inflexible social cognitive models that drive extreme views of others and negatively-skewed social attributions [3]. Specific hypotheses about social cognition and belief updating can be tested by estimating model parameters from behavioral tasks like the SVT.

*II.B. Models informed by known properties of circuits and specific cell types*
Experiments can also be designed to apply behavioral and fMRI data to computational models that are informed by knowledge about cell-type specific properties and circuit function. Increasing biological plausibility expands the number of junctures at which the facts of behavioral neuroscience can be leveraged in the model and ultimately exploited clinically. For example, models of reward and novelty-seeking behavior could emphasize the roles of dopamine neurotransmission in control of precision/uncertainty and sensory prediction errors as well as cholinergic regions of the brainstem and the noradrenergic locus coeruleus for encoding expected versus unexpected uncertainty (reviewed in [2]. Models based on predictive coding hold that cognition and perception can be understood in terms of prior predictions about inputs, discrepancies (prediction errors) between those prediction and actual inputs, the precision (or inverse variance) of those predictions and errors. Precision governs the weighting of the representation – towards the initial prediction (in which case the prediction error is ignored), or towards the error (garnering new learning). There is an assumption (and some data – [7]) that slower catecholaminergic neurotransmitters mediate this precision in neural hierarchies. For example, acetylcholine codes the precision of visual perceptual predictions, and oxytocin may code precision in the inferential hierarchy that governs social beliefs.

*II.C. Common tasks and models deployed across species*
Experiments can also be designed so that the same (or nearly-same) task is deployed in different model systems, for example, decision-making tasks can be performed by both patients and mouse models. One example that is relevant to BPD is in testing the effects of social rejection. In people, social rejection leads to alterations in subsequent social behavior [10] [11] and decreased pain sensitivity [12]. In rodents, adolescent social rejection produces effects on behavior and pain tolerance that mirror the results in human subjects [13]. An innovative paradigm also examined trust in the same rodents: animals were offered a novel food that they could smell on another rat's whiskers. Rats that underwent social rejection in adolescence were less trusting – they were less willing to eat that peer-introduced novel food [13]. Developing these parallel paradigms will allow meaningful cross talk between clinical and basic researchers, leveraging the precision and efficiency we can achieve in pre-clinical settings to investigate clinical hypotheses and novel treatments.

### III.    Approaches that may lend themselves to increased modelling in BPD
**III.A. Symptom provocation: an approach to examining fluctuating symptoms in the lab**

To study the problems that arise in social interaction, conducting experiments in live interactive settings may give a window to deficits that would be missed in static paradigms. Leonard Schilbach has coined the term "second-person neuroscience" to describe research that tests subjects in the closer contexts with study partners that shift mental processes from the distant "him" (third-person) to the more intimate "you" (second-person) [14]. In BPD, problems arise in social interaction, especially in the context of social stressors and perceived or actual interpersonal ruptures (reviewed in [15]). We outline several methods that have been used in BPD to engage interaction and social stress in experimental paradigms.

### III.A.1. Personalized narratives

Personalized narratives have also been used to temporarily stimulate stress by focusing on material relevant to each individual. This approach to experimental challenge sessions was pioneered in our department and used in multiple subject populations to investigate sometimes-fleeting experiences such as craving in substance-dependent patients [16, 17] and self-injurious impulses in people with BPD [18]. A variation on the stress narrative has recently been used in BPD to examine trust in the critical setting of an intimate relationship. Women with BPD, and controls without BPD, were invited into the lab to have videotaped conversations with their romantic partners. BPD patients experienced a greater loss of trust toward their romantic partner than did control subjects after the couples discussed personal and relationship threats, but not after they discussed a neutral topic [19]. We imagine that this paradigm might provide a probe of social dysfunction in BPD. Its effects may manifest in subsequent decision-making paradigms more amenable to computational modeling and the distillation of model parameters.

### III.A.2. Cyberball

Computer games offer an opportunity for social interaction with tight control of specific contingencies by the experimenter. In the social rejection paradigm Cyberball, players pass a ball back and forth on the screen [20] . The subject (who controls one of three players on the screen) believes she is playing with other people, but the other characters are controlled by the program. By varying the percentage of the time the ball is passed to the participant, feelings of social rejection can be evoked. In the most extreme case, the participant passes the ball to another player, and the other players only pass to each other, never returning the ball to the participant. This experience elicits sadness and anger in as few as six rounds of play [21], (meta-analysis in [20]). Individuals with BPD and healthy controls report similar emotions after playing Cyberball; the BPD players however report much more intense negative emotion [22]. Furthermore, people with BPD report feeling excluded not only during the rejection paradigm (the other players stop sharing the ball), but also during the fair paradigm (when all players receive the ball an equal number of times), and, of particular interest, the BPD players report feeling excluded despite having accurately perceived how often they got the ball [21]. Negative emotions are reduced, but not fully eliminated, when subjects with BPD receive the ball more times than any other player [22].

In non-psychiatric subjects, the insula is activated under conditions of unfairness and social rejection [23]. This discrepancy between perceived exclusion and actual inclusion may be a result of a failure to modulate activation of the insula and precuneus between different conditions among subjects with BPD.

### III.A.3. Virtual reality environments

Computer-based paradigms can take a step closer to live interaction with the use of virtual reality environments and avatars. For example, recent work in proxemics (the study of physical interpersonal behavior such as distance regulation, gaze direction, and posture) has assigned an avatar (virtual person in the simulated environment) to the subject whose behavior is controlled by the subject, then asked the subject to use her avatar to interact with other avatars (reviewed in [24]). This method allows the researchers to hold many variables constant as they tested the impact of specific cues, such as those that might reveal implicit biases. Of particular interest here, virtual environments and avatars can be used to suggest that a particular person is behaving is a specific expected or unexpected way. For example, one group claimed during an experiment that the second avatar was being controlled by the subject's romantic partner, and they tested the impact of attentive and rejecting behavior from the partner in a stress-provocation paradigm. These virtual reality contexts can facilitate collection of rich datasets to quantify complex behaviors such as interpersonal stance and approach/avoidance behavior.

### III.B. Quantifying social behavior
### III.B.1. Trust game data
Neuro-economic games can be powerful tools to help dissect specific aspects of social behavior. One task, the Trust game (reviewed in [25],[26]) is played by two individuals (or a person and a computer) with different roles: investor and trustee. At the beginning of the game, the investor receives a small sum of money. The investor can choose to transfer any portion of the money to the trustee. The trustee receives triple the transferred amount, and can then choose to return any portion (none up to the whole amount) back to the investor. The game is typically played over 5-10 rounds. To maximize reward, the investor and trustee must trust each another enough to transfer money. Investors with BPD transfer less money to trustees than do healthy or depressed control subjects, and are less optimistic about how much money will be transferred back [27]. Interestingly, these differences disappear when people with BPD are told that money transferred back will be randomly determined by a computer and not by a human player [27]. Trustees with BPD also have difficulty cooperating with the other player, tending to receive smaller investments from partners as the game progresses [28]. While healthy controls have been found to have increased activation in the bilateral anterior insula when investments were smaller, the insula response in subjects with BPD did not correlate with investment level [28]. Also, in healthy players, a common response to small investments is coaxing: returning a large percentage to the investor, signaling trustworthiness. Trustees with BPD are less likely than healthy control trustees to employ this strategy, which may explain their partners' decreased investments over the course of the game [28]. BPD subjects were also less able to use facial cues to update their assessment of trustee intent [29].

### III.B.2. Learning data
Several small studies have examined learning in BPD. Berlin et al compared probabilistic reversal learning in 19 subjects with BPD, 23 with orbito-frontal (OFC) lesions, 20 with pre-frontal lesions not including OFC, and 39 healthy control adults [30]. Only OFC lesion subjects differed from control in the learning task measures (overall learning, reversal learning, and punishment sensitivity), though BPD and OFC subjects both had increased impulsivity and problem behaviors. In another small study of 10 students endorsing BPD symptoms on a self-report scale versus 128 healthy control students, there was no difference in reversal learning [31]. Barker et al tested [32] also confirmed this finding in 20 BPD subjects versus 21 healthy control subjects, and by contrast, found between-group differences in reversal learning (only for all the trials analyzed together) as well as small significant differences in intra-dimensional and extra-dimensional shifting.

Paret et al. tested reversal learning in a task with social stimuli. In this setting, subject arousal and degree of dissociation correlated with difficulty with learning (acquisition), whereas self-reported BPD symptom intensity correlated with difficulty with reversal learning [33]. Multiple groups have examined working memory in groups of 20-40 BPD subjects versus healthy control and found no difference [30], [34], suggesting that the observations of impaired counterfactual social inference are not merely a reflection of broader cognitive impairment in BPD, but rather specific neurocognitive deficits.

III.C. Pulling in data from the real world

III.C.1. Mechanical Turk: widely disseminated research data collection
Crowdsourcing sites are an effective way to recruit large numbers of participants into a study at low cost. The most prominent of these sites is Amazon's Mechanical Turk (mTurk).  mTurk was designed to allow businesses to outsource tasks that would ideally be automated, but are currently better done by humans, like tagging photos or looking for damage in images of roads. The tasks are completed by ordinary individuals with Amazon accounts, who are paid through the site.  mTurk has been validated as a reliable research tool [35]. US mTurk workers are more diverse than college student samples, though they are still  younger, more liberal, and more European/Asian than the general population [36], [37]. Results of mTurk studies reveal self-report scale reliability, task to task consistency, and worker attentiveness that are indistinguishable from other samples [38] [39], [40].  Subjects can be re-contacted for a series of studies. The site lends itself to a wide variety of research methods: surveys, group games, and even eye-tracking tasks using the workers' webcams have been done [41], [42].  In sum, mTurk represents a powerful pipeline for recruitment of large numbers of subjects outside traditional research settings at low cost. mTurk may be a good venue for testing research hypotheses in people with BPD and BPD features, especially those that do not present for clinical attention.  As has been done in other settings, researchers could collect behavioral data after stratifying subjects by response to self-report scales to facilitate dimensional analyses.

III.C.2. Ecologic momentary assessment (EMA)
Real-world and real-time data collection can include EMA for a range of indicators of interest, for example GPS coordinates versus known locations of interest (home, work, friends' homes, emergency room, liquor store), biophysical data (cardiovascular parameters, exercise, skin conductance, gestural data), speech recordings, as well as frequent brief symptom self-reports [43].  In BPD, EMA is particularly relevant given the frequency of symptom fluctuations, and it has been used to clarify the relationship of personality traits to current-state experiences and future functioning [44] [45]. These larger data sets with finer grained time courses may allow us to study cyclical changes such as menstrual cycle and other fluctuating and part-time contributors to symptom fluctuation.

III.C.3. Language
Analysis of individual and group speech and writing samples can be undertaken with both real-world and laboratory-acquired language samples.  We and others have identified language features that mark psychological states and traits. For example, self-referentiality in writing increases with authors' depression and suicidality, and separates patients (people with a diagnosed illness) from non-patients [46].  Computational models based on word-use patterns can predict which writers have psychosis [47],[48],[49], or will progress to psychosis [50].

Study of language and conversation in interaction may be worth particular attention in BPD. Jamie Pennebaker's group has identified patterns of pronoun use during speed dates that correlate to later relationship trajectory [51]. We expect that studies engaging methods such as the couple conversations referenced above [19] may be ideal settings to identify language features that fluctuate with interpersonal instability and other symptoms in BPD. Sound and content of speech could offer non-invasive markers of current illness status, but given the many complex features, good models together with large corpori will be important to meaningfully relate language features to clinical concerns.

III.D. Neural circuits
Research in the structural and functional neuro-anatomy of BPD has yielded a few key results, and some potentially interesting but not well-replicated findings.

Negative attribution bias in BPD has been linked to hyper-active amygdala (at rest also in response to stimuli). A recent meta-analysis of neural response to emotional stimuli confirms greater left amygdala activation as well as decreased dorsolateral prefrontal cortex (dlPFC) response in people with BPD versus controls [52].

Efforts to describe differences in brain networks in BPD are starting to be reported. A meta-analysis found 7 reports of resting state data, with overall significant increase in activity in the midline and dorsal structures of the default mode network [53]. Xu *et al.* modelled the topology of networks in resting state data from 20 people with BPD and 10 controls, finding differences in both local and global connectivity patterns that correlate with symptom measures [54].

The next step will be to test neural responses in the context of increasingly interactive and attachment-engaging settings. In a non-social monetary incentive delay task, people with BPD show reduced striatal activation vs. control in response to both anticipated reward and anticipated loss [55]. However, this pattern of findings is recapitulated across diagnostic groups, and may reflect the stigmatizing effects of having a mental illness, the increased proclivity toward substance abuse or the socioeconomic consequences of being mentally ill. We suggest that focusing on processes more directly relatable to individuals' symptoms, may be a more fruitful line of enquiry. We discussed above the differential insula signal in a brief trust game with a stranger [28]. Looming photographs of faces with negative emotions differentially activate the amygdala and somatosensory cortex in BPD versus control [56]. Future work is needed to test how these responses may differ with tasks that better simulate live interaction, when playing under social stress, and when playing with others who matter to the subject. Games like the SVT we described above may be useful – neural signals have been carefully described in healthy subjects in response to surprises about the social and non-social rewards [8]. We hypothesize that people with BPD may differ in this neural mechanism for belief updating, a possible explanation for rigid and extreme social representations [3]. We expect that computational modelling will not only allow researchers to test this kind of hypothesis, but moreover to be encouraged to develop more specific and testable hypotheses about social learning in BPD.

III.E. Molecular and genetic markers

III.E.1. Oxytocin
Neuropeptides, including oxytocin and vasopressin, have been implicated in normal social cognition, as so garnered particular attention as candidates for molecular underpinnings of problems in disorders with prominent social symptoms such as Austism Spectrum Disorders and Borderline Personality Disorder [57] [10] . In healthy animal and human models, oxytocin has been found to modulate social perception, social learning, and pain [58]. However, the exogenous administration of oxytocin to people with BPD yielded surprising results: though in some settings, oxytocin decreased both self-reported stress and cortisol levels, behavior in the trust game was less trusting and less cooperative (in BPD but not in controls). This finding conflicts with a simple explanation of oxytocin as a pro-affiliation hormone, but opens intriguing questions about how the social brain may differ in BPD.

III.E.2. Other molecular players

A wide range of other molecules have been studied in BPD with mostly conflicting results, and some small studies that bear replication (reviewed in [59]). Of particular note, data about resting and provoked Hypothalamic-Pituitary-Adrenal (HPA) axis function are many and conflicting. Small studies suggest decreases in peripheral BDNF and increases in peripheral VEGF, raising questions about the role of neutrotrophic support to maintain adult plasticity and neuronal health. Serotonin receptor function has been examined, and two PET studies suggest increased binding potential at 5HT2A receptors. Data also suggest sex differences in the serotonergic system in BPD.

The role of epigenetics is now well-established in propagating the effects of childhood trauma [60] [61] [62]. Given the increased risk for BPD among people with trauma in early childhood, researchers have begun to examine epigenetic markers and polymorphisms in this population [63] [64].

III.F. Models of self and self-other interaction
We and others have argued elsewhere that computational models are critical to testing specific predictions about how people understand themselves, and themselves with respect to others in the social arena. One relevant framework is predictive coding, which argues that people not only respond to perceptual data and update their worldviews in response, but that top-down expectations actively modulate the processing of perceptions such that expected states have a profound impact on the states that are ultimately experienced. This is relevant to a variety of contexts in mental illness, but has been not much applied to BPD yet.

III.F.1 Testing hypotheses about self-other psychology in BPD
Several current hypotheses about mental functioning in BPD could lend themselves to validation using a model-based method. For example, Marsha Linehan and others suggest in the biosocial model that underlies DBT that people with BPD are quicker to make negative attributions in an upsetting social setting, and slower to normalize those attributions [65]. As discussed above, we can use predictive models to test learning rates as various contingencies change (likelihood or stability of reward, value of the reward). Future studies may be able to engage a more BPD-relevant approach with social instead of financial reward metrics. Psychodynamic theories suggest that people with BPD have rigid extreme internal representations of others (unintegrated objects) [66] [67] – we can model these as predictions (prior estimates) about others' intentions and others' future behavior, and test for differences in model parameters in BPD and control.

III.F.2. Testing hypotheses about the role of the peripheral body in self-concept in BPD
Sense of self involves a complex interplay of cognitive representations and peripheral data. Disturbed sense of self is a key feature of BPD, and includes symptoms that are explicitly physical. Indeed, people with BPD evince diminished interoceptive signals than do controls: amplitude of heartbeat-evoked potentials, a neural response to felt heartbeats, were lower in amplitude in BPD [68].

In other settings, tests of illusory body ownership have been used to examine how easily subjects lose a sense of their own physical boundaries. For example, in the rubber hand illusion, subjects watch a rubber hand get stroked with a paintbrush while they feel their own hidden hand also stroked. This experience can lead to the sense that the rubberhand is one's own hand, and illusion-susceptibility correlates with anterior insula activity (reviewed in [69]). Illusion-susceptibility varies with psychopathology, for example schizophrenia, and also experimental induction of psychotomimetic experience in healthy control subjects. Both increase susceptibility to the illusion [70] [71]. Furthermore, experience with this and related illusions varies with specific symptoms, such as dissociation [72]. Anil Seth and others have argued that the anterior insula is a key nexus for the prediction error driven inferences that guide perceptions of bodily ownership and agency [73-75]. Others highlight the parietal cortex as a key locus for the illusion, and the sense of agency over body and limbs more broadly [76]. The illusory body tasks have also now been extended to animal models with the report that mice are susceptible to a "rubber tail illusion" [77]. Not only will illusory-body tasks in BPD help to arbitrate questions about self-concept. Might people with BPD be less susceptible to the illusion as they are less sensitive to physical cues (as in pain studies)? Or might they be more susceptible to the illusion, consistent with the confused boundaries of identity diffusion? Could we test underlying biology in the adolescent social rejection mouse model? Or track progress in psychotherapies that aim to firm up understandings of self and other, such as Transference-Focused Psychotherapy, by following rubber hand illusion response?

III.F.3. Perceptual aberrations

People with BPD can experience frank perceptual aberrations, sometimes termed "micropsychotic" symptoms. These experiences can include hallucinated voices. In BPD, voices often arise under the stress of relationship threat. The schizophrenia researcher Ralph Hoffman conceptualized the etiology of voice hearing through a mechanism he termed "social de-afferentation", in which the brain responds to social isolation by developing a markedly increased sensitivity to potential social cues, and so misperceives auditory cues as social input, in particular, voices of other people talking.

Computational modelling has recently been applied to this problem. Al Powers and colleagues have recently tested a predictive-coding-based hypothesis that hallucinations arise from over-weighted top-down expectations during perception (paper under review). They were able to engender hallucination-like experiences using only classical conditioning and found that individuals with hallucinations (regardless of the presence or absence of other psychotic symptoms and dysfunction) were markedly more susceptible to these conditioned hallucinations, which were correlated with activity in a network of brain regions that closely mirrored those typically seen in symptom-capture-based approaches to hallucinations. Lastly, formal modeling of perception during conditioned hallucinations using the Hierarchical Gaussian Filter (HGF) was used to parse this network into regions underlying specific computational functions and to identify differences between groups. Of note, in their dataset, people with schizophrenia are significantly more likely to meet criteria for BPD on the SCID-2 self-report questionnaire if they hear voices than if they do not [78]. There are also some tempting genetic targets, for example the forkhead box P2 gene (FOXP2) has been found to interact with childhood maltreatment (a risk factor for BPD) to confer risk for auditory verbal hallucinations [79]. Future studies modelling aberrant perceptions in BPD may especially benefit from examining interactions between early adversity, genetics, and neural function.

IV. **Conclusions**

In this paper, we have reviewed the key elements of the computational psychiatry approach, focusing on examples in social neuroscience. We discussed the methods that may be particularly relevant to developing a computational psychiatry of BPD. These start with interactive social paradigms, both live and computer based. We reviewed methods for amplifying relevant social experiences with symptom provocation by social rejection tasks (e.g. Cyberball) and autobiographical memory scripts. We discussed the importance of quantifiable social tasks and the useful intersection of interaction and easy quantification in economic games. Quantification of social behavior may be even more relevant to patient experience when it can occur closer to the real world: we discussed crowd-sourced samples and real-time measurements toward this end. We hold that efforts should be made to collect biological data together with behavioral outcomes, and have suggested that this discussion should also engage molecular pathways and neural circuits. Lastly, we considered the physical self and the potential role of embodied cognition in BPD.

In summary, computational approaches to behavioral and neural data are already rapidly pushing the boundaries in BPD research to yield increased cross-talk among basic scientists, neuroimagers, and clinical scientists. Optimism about more relevant, mechanistic, and treatment-driving results is well-founded. Collaboration with computational modellers will be a cornerstone of progress as we move forward in the field.

## References:

Papers of particular interest, published recently, have been highlighted as:

• Of importance

•• Of major importance

## Of importance:

**Reference 50 (Bedi et al. 2015):** This paper is a very nice example of using behavioral data (here, language from people with first-episode schizophrenia) in a complex computational framework to predict prognosis.

**Reference 52 (Schulze et al. 2016):** This meta-analysis is a well-architected and clearly written update on the state of neuroimaging in BPD focusing on functional responses to negative stimuli.

**Reference 74 (Seth AK 2013):** This commentary is a well-written and thoughtful discussion of the neural basis of embodied cognition, a field particularly relevant to BPD given the prominent somatic symptoms and (related) disruptions to sense of self.

# Of major importance:

**Reference 1 (Wang et al. 2014):** This review paper describes the computational psychiatry approach at multiple levels from behavior to molecules. Figures are especially helpful to clarify concepts.

**Reference 9 (Sevgi et al. 2015):** This paper describes a computer-based game that requires subjects to weigh social and non-social cues to make decisions. The task is applied to people with varying levels of autistic traits. A computational learning model, the Hierarchical Gaussian Filter, is used to measure differences in social and non-social learning behavior.

**Reference 13 (Schneider et al. 2016):** This paper describes a rat model of early-life social rejection that mimics many of the adult behaviors of people with BPD, such as decreased trust of peers and decreased pain sensitivity. This model holds promise for translational and reverse-translational experimental pipelines in BPD.

**Reference 14 (Schilbach et al. 2013):** This review and commentary puts forth the position that we are doing social neuroscience wrong: instead of studying people in isolation (first-person) or people with dry unmoving social cues like faces (third-person), we should instead study people in interactive contexts to learn the most about their social cognition. This will be the second-person neuroscience.

**Reference 28 (King-Casas et al. 2008 – included bc there have been no definitive follow-ups more recently and this is a key paper to this argument):** Behavioral and neuroimaging results are presented from a study of people with BPD playing the Trust Game. Results include less cooperative behaviors, like coaxing a defecting partner back to play, and diminished ability to detect fair offers.

**Reference 68 (Muller et al. 2015):** The researchers examined the neural signal (in EEG) that reports on individuals' heartbeats. Consistent with altered sense of physical self in BPD, they found decreased amplitude in BPD and an intermediate phenotype (between BPD and control) for people who have recovered from BPD. This suggests that brain signals may normalize as symptoms decrease.